\documentclass[preprint2]{aastex}
\usepackage{amsmath}

\def\lsim{\mathrel{\rlap{\lower4pt\hbox{\hskip1pt$\sim$}}
    \raise1pt\hbox{$<$}}}                % less than or approx. symbol
\def\gsim{\mathrel{\rlap{\lower4pt\hbox{\hskip1pt$\sim$}}
    \raise1pt\hbox{$>$}}}                % greater than or approx. symbol
\slugcomment{}

\shorttitle{Long duration flare in TWA 11B}
\shortauthors{L\'opez-Santiago et al.}

\begin{document}


%\title{Extremely long duration flare rise-phase in TWA 11B: 
%large loop structures in a \textit{not so young} M star}

\title{A detailed study of the rise phase of a long duration X-ray flare 
in the young star TWA 11B}


\author{J. L\'opez-Santiago\altaffilmark{1} and
             I. Crespo-Chac\'on\altaffilmark{1}}
\affil{\altaffilmark{1}Departamento de Astrof\'{\i}sica y 
      CC. de la Atm\'osfera, Universidad Complutense de Madrid, 
      E-28040 Madrid, Spain \vspace{0.2cm}}

\and 

\author{\vspace{-0.4cm}G. Micela\altaffilmark{2}}
\affil{\altaffilmark{2}INAF - Osservatorio Astronomico di Palermo
                                 Piazza Parlamento 1, I-90134 Palermo, Italy}

\and

\author{\vspace{-0.4cm}F. Reale\altaffilmark{3, 2}}
\affil{\altaffilmark{3}Dipartimento di Scienze Fisiche e Astronomiche, 
                                   Sezione di Astronomia, Universit\`a di Palermo, 
                                   Piazza Parlamento 1, I-90134 Palermo, Italy}

%\email{aastex-help@aas.org}


%\altaffiltext{1}{Visiting Astronomer, Cerro Tololo Inter-American Observatory.
%CTIO is operated by AURA, Inc.\ under contract to the National Science
%\altaffiltext{2}{Society of Fellows, Harvard University.}
%\altaffiltext{3}{present address: Center for Astrophysics,
%    60 Garden Street, Cambridge, MA 02138}
%\altaffiltext{4}{Visiting Programmer, Space Telescope Science Institute}
%\altaffiltext{5}{Patron, Alonso's Bar and Grill}


\begin{abstract}
%In this paper, 
%we analyze a long duration flare observed in a serendipitous 
%XMM-Newton detection of the M star CD-39 7717B (TWA 11B), member
%of the young stellar association TW Hya. Only the rise phase (with
%a duration of $\sim 35$ ks) and possibly the flare peak were observed. 
%%During the exposure, the X-ray flux increased of a factor of $\sim 4$.
%We took advantage of the high count-rate of the X-ray source to carry out
%a detailed analysis of its spectrum during the whole exposure, what 
%allowed us to compare the results given by the single-loop and two-ribbon 
%models for the flare parameters. After a careful analysis, 
%%Taking the light curve 
%%and the evolution of the hardness ratio into account, 
%we interpret the rise
%phase as resulting from the ignition of a first group of loops (event A)
%which triggered a subsequent two-ribbon flare (event B). Thus, event A 
%was analyzed using the single-loop model. For event B, the diagnostic 
%method for two-ribbon flares was applied.
%Loop semi-lengths of about $2.5 - 3$ stellar radii were obtained. Such
%large structures had been previously observed in very young stellar 
%objects ($\sim 1-3$~Myr). This is the first time that they have been 
%inferred in a more evolved star. Finally, The fluorescent iron
%emission line at  6.4 keV was detected during this event. As far as
%we are concerned, since TWA 11B seems to have no disk, this is only
%the third detection of Fe photospheric fluorescence.
We analyzed a long duration flare observed in a serendipitous 
XMM-Newton detection of the M star CD-39 7717B (TWA 11B), member
of the young stellar association TW Hya ($\sim 8$~Myr). Only the rise 
phase (with
a duration of $\sim 35$ ks) and possibly the flare peak were observed. 
We took advantage of the high count-rate of the X-ray source to carry out
a detailed analysis of its spectrum during the whole exposure. 
%This allowed us to compare the flare parameters resultant of using both a 
%single-loop and a two-ribbon model. 
%
After a careful analysis, we interpreted the rise
phase as resulting from the ignition of a first group of loops (event A)
which triggered a subsequent two-ribbon flare (event B). Event A 
was analyzed using a single-loop model, while a two-ribbon model 
was applied for event B.
Loop semi-lengths of  $\sim 4 R_\mathrm{*}$ were obtained. 
Such large structures had been previously observed in very young stellar 
objects ($\sim 1-4$~Myr). This is the first time that they have been 
inferred in a slightly more evolved star.  
The fluorescent iron emission line at 6.4~keV was detected during 
event B. Since TWA 11B seems to have 
no disk, the most plausible explanation found for its presence in the X-ray 
spectrum of this star is collisional- or photo-ionization.
As far as we are concerned, this is only the third clear detection of Fe 
photospheric fluorescence in stars other than the Sun.

\end{abstract}





\section{Introduction}
\label{sec:int}

Among all the processes manifested in stellar coronae, 
flares are the most energetic ones. They are supposed to be the result of the 
energy release from magnetic field reconnection in the lower corona 
\citep[e.g.][]{kop76}. 
%The energy is released somewhere in the tubes.
As a consequence, electrons and ions in the reconnection region 
are accelerated downwards, along the magnetic field 
lines, toward lower atmospheric layers. When they reach 
the upper chromosphere, the local gas 
is heated and evaporated into the new-formed magnetic loops. Thus, the density and 
temperature of these loops increase, causing intense emission of soft 
($<10$~keV) X-rays.

Stellar flares have been observed in almost all the \mbox{H-R} diagram \citep[see][for a 
review]{vai81}. However, they are more frequent in late-type stars, where
they present a large variety of sizes and durations. 
In late-K and M dwarfs \citep[the so-called UV Ceti-type stars; see][for 
a description of their main properties]{pet91}, moderate flares are
frequently observed. During such events, the X-ray flux usually 
increases by a factor of  \mbox{$2 - 4$} \citep[e.g.][]{rob05}. 
Giant flares, in which the X-ray flux increases from dozens to hundreds times the 
quiescent state value, have been detected in some M dwarfs such as \object{EV~Lac} 
\citep{fav00}, \object{EQ~Peg} \citep{kat02}, and \object{Prox~Cen} \citep{gue04}.
The duration of those flares
are of the order of a few kiloseconds, even in the more energetic ones. 
For instance, the giant flare observed in \object{EV~Lac} lasted 
$\sim 5$~ks. Long-duration flares are more common in pre-main sequence stars
than in the UV Ceti-type ones. For example, 
during the 13 days observing run of the ORION Nebula Complex by \textit{Chandra}, 
at least 19 flares with durations above half a day were detected \citep[][]{fav05}. 
Long-duration flares were also observed in the Taurus star-forming complex
\citep[][]{fra07,ste07}. The longer duration of these events is usually attributed to 
the presence of larger coronal structures in young stars.

%Giant flares in which the X-ray flux increases dozens to hundreds times the 
%quiescent state value have been detected in some M dwarfs such as \object{EV~Lac} 
%\citep{fav00}, \object{EQ~Peg} \citep{kat02}, and \object{Prox~Cen} \citep{gue04}. 
%Nevertheless, in late-K and M dwarfs \citep[the so-called UV-Ceti-type stars; see][for 
%a description of their main properties]{pet91}, moderated flares are far more
%frequently observed. During such events, the X-ray flux usually 
%increases by a factor of  \mbox{$2 - 4$} \citep[e.g.][]{rob05}. The duration of those flares
%are of the order of several kiloseconds, even in the more energetic ones. 
%For instance, the giant flare observed in \object{EV~Lac} had a total duration 
%of $\sim 5$~ks. 
%%
%On the other hand, 
%long-duration flares have been more typically observed in pre-main sequence stars. 
%During the 13 days observing run of the ORION Nebula Complex by \textit{Chandra}, 
%at least 19 flares with durations above half a day were detected \citep[see][]{fav05}. 
%In the Taurus star-forming complex, other long-duration flares were also observed 
%\citep[see][]{fra07,ste07}. The longer duration of these flares is usually attributed to 
%the presence of longer coronal structures in young stars.

The \textit{XMM-Newton} and \textit{Chandra} missions have 
contributed enormously to the understanding of the processes involved in the X-ray 
emission of late-type stars. In particular, the improved temporal and spectral resolution, 
together with the development of theoretical models \citep[e.g.][]{kop84,pol88,ser91,
gue99,rea04,rea07}, have provided us with powerful tools for investigating 
%the characteristics of 
coronal flares. 
Detailed diagnostics of X-ray flares have been carried out by different authors. 
\citet{fav05} determined general properties of flaring loops, in terms of temperature
and semi-length, for young stellar objects in the Orion Nebula Complex (ONC). 
>From their analysis, the authors inferred loop semi-lengths comparable to 
the stellar radius in some cases. They speculated that these large structures are 
connected with the proto-planetary disk and are, in fact, the same structures that
channel the plasma producing accretion. A similar 
work was done for the Taurus star-forming region \citep{fra07}, where 
the authors remarked that some coronal loops extended up to 
a distance comparable with the stellar radius 
\citep[see also][]{gia04}.
%In this case, no indication of some
%%the 
%connection existing between long loops and stellar disks was detected 
%\citep[see][although]{gia04}. 
%
Studies for 
%with 
more evolved stars were done by, e.g., \citet{rea04}, \citet{cre07}, and \citet{tes07}, 
who found 
%the heated plasma confined in magnetic tubes to have 
flaring loops with semi-lengths of $\sim 0.2 - 0.5$~$R_\star$ 
(i.e., a relatively compact flaring corona).
%. This implies a relatively compact flaring corona.
%%for these stars. 

In the works presented in the previous paragraph, the diagnostic of X-ray flares 
was done using the 
procedure described by \citet{rea97, rea04} to model the decay phase, 
which assumes the flare to be produced in a single loop 
where heating does not entirely drive the flare decay.
\citet{rea07} extended this  method to the rise phase and compared 
the parameters determined in this way with those obtained by analyzing the decay 
phase for three stellar flaring loops \citep[see also][]{pan97}, 
finding a good agreement between them.
On the other hand, the solar two-ribbon model developed by \citet{kop84}, 
or its stellar version \citep{pol88}, should be used
when heating totally drives the flare evolution \citep{rea02,rea03}. 
In this model, 
the reconnection energy is supposed to be dissipated immediately after being released.
\citet{kop84} assumed that only a fraction of the magnetic energy released by the
reconnection process is used to supply the thermal energy of the newly formed
flare loops. \citet{pol88} assumed that a factor of $10~\%$ of this
thermal energy escapes into the X-ray regime,
%took into account conductive losses 
%%(which transfer thermal energy from the X-ray loops to other flare regions, where it is
%%eventually radiated away in the form of UV lines, H$\alpha$, etc.)
%by assuming that the radiative losses observed in the X-ray band
%are a factor $\approx 10~\%$ of the magnetic energy released by the reconnection
%process, 
as suggested by detailed studies of solar flares \citep{can80}.
We also refer the reader to \citet{gue99}, who included time-dependent
conductive and radiative losses in the X-ray corona self-consistently.

In this work, we analyze an XMM-Newton serendipitous observation of the young 
M-type star TWA 11B in which the X-ray emission suffered a continuous increase 
during approximately 35\,ks. The duration and statistics of the observed rise  
have allowed us: (i) to carry out a detailed spectral time-analysis; and 
(ii) to derive properties of the star's magnetic configuration 
by using both the single-loop and the two-ribbon flare models
described above.
% in the different parts detected in the rise phase. 

%________________________________________________ Figure
\begin{figure}[!t]
%\label{fig1}
\includegraphics[width=7.5cm,clip=true]{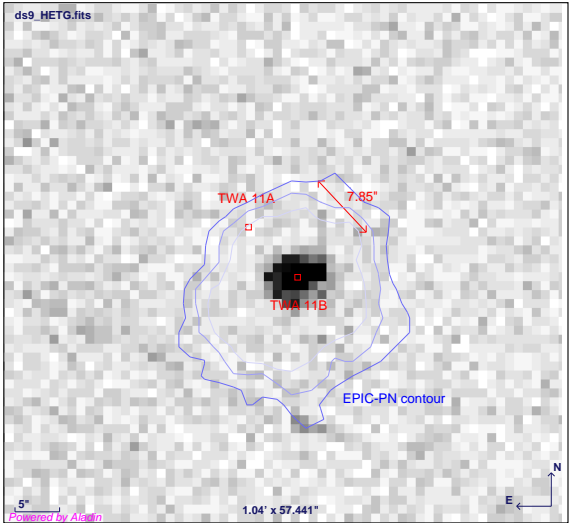}
\caption{Chandra HETG serendipitous detection of TWA 11B. The 
position of the optical counterparts of TWA 11A and TWA 11B are marked. 
Contours of the PSF observed in the XMM-Newton EPIC exposure are 
overplotted.}
\label{fig0}
\end{figure}
%______________________________________________________

\section{Observation and data treatment}
\label{sec:observations}

The XMM-Newton observation (ID 0006220201) was perfomed in the revolution 
197, between 2001 January 4 and 5, for a total duration of $43$\,ks.
The field is centred on the coordinates
$\alpha_\mathrm{2000} = 12^\mathrm{h}35^\mathrm{m}34^\mathrm{s}$ and
$\delta_\mathrm{2000} = -39^\mathrm{\circ}54^\mathrm{'}55^\mathrm{''}$ 
(the main target being the Seyfert 2 galaxy \objectname{NGC 4507}).
%($\alpha_\mathrm{2000} = 12^\mathrm{h}35^\mathrm{m}34^\mathrm{s}$,
%$\delta_\mathrm{2000} = -39^\mathrm{\circ}54^\mathrm{'}55^\mathrm{''}$).
The EPIC cameras were operated in Full Frame mode using the Thick filter.
Our target is situated 5.18\,arcmin at the North-East of the central source, in 
the field of view of both the MOS and PN detectors.
{The EPIC-XMM source coincides  with the position of the optical 
counterpart of TWA 11B. The primary 
star, TWA 11A (an A0-type star), is situated at $\sim 8$ arcsec from TWA 11B. 
Although none X-ray emission is expected from an A0 star, we investigated the 
possibility that the primary star produced some X-rays that could affect our results.
We studied a 
serendipitous detection of HETG Chandra of the source (Observation ID 2150). 
In this 140 ks exposure Chandra observation, the X-ray source is clearly identified 
with TWA 11B (see Fig.~\ref{fig0}), while none X-ray emission coming from TWA 11A 
is detected. Therefore, we are confident that the X-ray emission detected in the 
XMM-Newton observation comes only from the M star TWA 11B.}

The data reduction followed the standard operating procedure. We used the 
version 7.1.0 of the XMM-Newton Science Analysis System (SAS) to derive a 
table of calibrated events in the energy range 0.3 -- 10.0\,keV. To extract the 
events, we chose a radius of 30\,arcsec, which is slightly larger than the
3$\sigma$-level of the source's PSF. This assures us to lose less than 1\%
of the counts from the source. Different filters were applied to eliminate bad
events and noise. Note that the observation was neither
%not
affected by pile-up
%or 
nor by high flaring background periods.

\begin{figure}[!t]
%\label{fig1}
\includegraphics[bb= 53 355 395 587,width=8.0cm,clip=true]{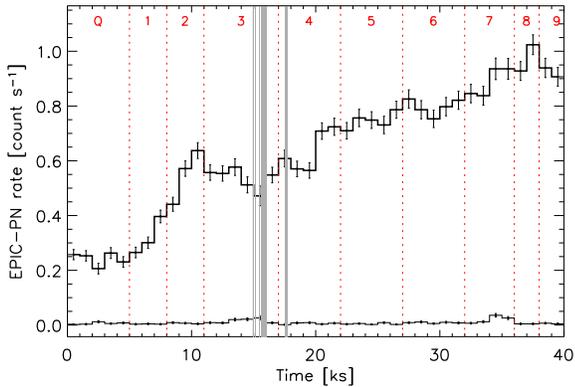}
\caption{EPIC-PN light curve of TWA 11B in the energy range $0.3 - 10.0$ keV.
The curve was binned to a 1\,ks time resolution. The event list was
corrected of bad events and noise. The exposure was also corrected of
live-time and good-time intervals. The gray segments mark the periods in
which the intrument was turned off. The continuous line at the bottom is the
background light curve. The time blocks used in the spectral analysis are plotted as 
dashed vertical lines.}
\label{fig1}
\end{figure}

The X-ray light curve (Fig.~\ref{fig1}) showed a total increase in the star's 
count-rate of a factor of $4.2$ from the lower level ($\approx 0.25$ 
counts/s) to the maximum observed emission. 
Similar relative increases in flux were previously reported in
flares from other stars \citep[e.g.][]{rob05}.
%The latter took place $\approx 3$ ks before the end of the observation. 
A first increase in the light curve is observed 
$5$\,ks after the beginning of the observation, reaching a local maximum only 
$5$\,ks later. After a brief decrease, the count-rate continues increasing 
until reaching the global maximum in the light curve
%at approximately 
(this happened 37\,ks after 
the beginning and only 3\,ks before the end of the observation).
The total duration of the enhancement is 
$32$\,ks ($\approx 9$ hours\footnote{{The rotation period of TWA 11B is not 
known. \citet{sch07} measured a projected rotational velocity 
$v \sin i = 12.11 \pm 0.93$ km\,s$^{-1}$, what leads to an upper limit in the rotational 
period of 2.7 days. Thus, the observation cover, at least, 17\% of the rotational period 
of the star. Nevertheless, no indication of occultation of the flaring 
region is observed.}}).
Note that the observed maximum may or may not be the flare peak.
In the latter case, the duration of the rise phase would be even
longer.

\begin{figure}[!t]
\includegraphics[bb= 50 5 480 324, width=8.0cm,clip=true]{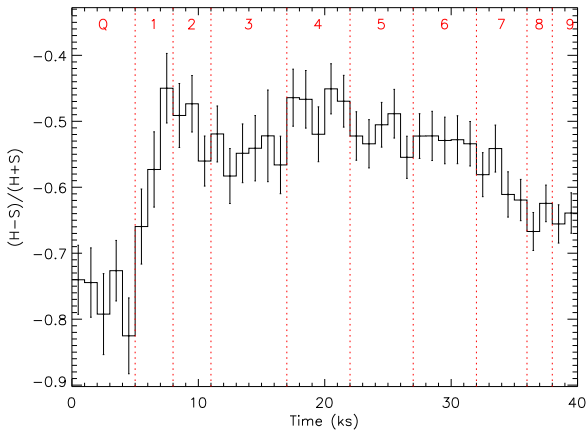}
\caption{Hardness ratio evolution of TWA 11B during the observation
with EPIC-PN. The time blocks used in the spectral analysis are plotted
as in Fig.~\ref{fig1} for clarity.}
\label{fig2}
\end{figure}

In Fig.~\ref{fig2} we plot the evolution of the hardness ratio, which is a tracer
of the temperature evolution, during the observation. Here,
the soft energy band is defined as the range 0.3 -- 0.8 keV
and the hard energy band as 0.8 -- 4.5 keV. Fig.~\ref{fig2} shows that the mean coronal
temperature
%during the observation
reached a maximum at $t \sim 7$ ks,
maintained high with significant fluctuations during approximately the next 27\,ks of
exposure, and then it began to decrease gradually.

%________________________________________________ Figure
%______________________________________________________

%________________________________________________ Figure
%______________________________________________________

%In terms of relative increase in flux, similar values have been 
%previously observed 
%%in flares on UV Ceti-type stars 
%\citep[e.g.][]{rob05}. 
%However, to date, so long rise phases have been seen only 
%in T Tauri stars 
%\citep{fav05, gia06, fra07}. In such stars, the relative increase in flux 
%ranges from a factor 3 \citep[V410 Tau;][]{fra07} to a factor 10 
%\citep[V892 Tau;][]{gia04} with rise-phase times from 8 to 55 ks. 
%TWA 11B is an M2.5 star with an age of $\sim 8$ Myr (estimated from its 
%membership in the TW Hya Association). This is clearly out of the range of 
%age of T Tauris but, nevertheless, during the observation it seems
%to have had a flare event very similar to those observed in some T Tauri stars. 

%From the light curve and the hardness ratio evolution, we interpret the observed 
%rise phase as resulting from the ignition of a first loop (or a group of loops 
%dominated by one of them) which triggered a subsequent two-ribbon flare. 
Overall, the rise phase is unusually long compared with those observed in other
stars \citep[e.g.][]{pan97,gue99,rea02,rob05,rea07,cre07}. 
To date, so long rise phases have been observed only in some T Tauri stars 
\citep{fav05, gia06, fra07}, {some of them with accretion disks, and several
RS~CVn systems \citep{tes07,nor07}.} 
In such stars, the relative increase in flux 
ranges from a factor 3 \citep[V410 Tau;][]{fra07} to a factor 10 
\citep[V892 Tau;][]{gia04} with rise-phase times from 8 to 55 ks. 
TWA 11B is a {weak-line T Tauri} M2.5 star with an age of 
$\sim 8$ Myr (estimated from its membership in the TW Hya Association). 
%This is clearly out of the range of 
%age of T Tauris but, nevertheless, during the observation it seems
%to have had a flare event very similar to those observed in some T Tauri stars. 
{While it shows no signatures of an accretion disk \citep[near-infrared color excess 
or strong H$\alpha$ emission;][]{sta95}, it is still contracting. Thus, 
physical conditions in its atmosphere should be more similar to those 
of sub-giant stars than of main-sequence ones. This may be the reason why 
long duration rise phase flares are observed in both T Tauri stars and RS CVn
systems.}

\begin{figure}[!t]
\includegraphics[width=8.0cm]{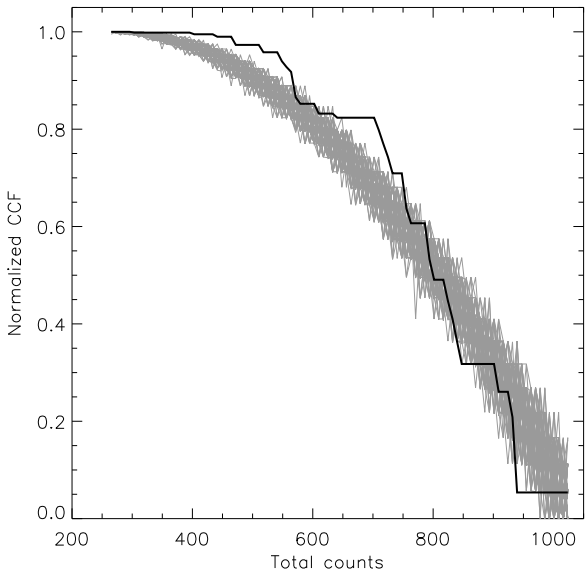}
\caption{Normalized cumulative distribution of counts in the light curve of TWA 11B
for $t \ge 5$\,ks (continuous line). The shadowed region contains cumulative curves 
of 200 simulated constantly increasing light curves (also for $t \ge 5$\,ks) with 
Poissonian noise.}
\label{fig2.1}
\end{figure}

A more detailed inspection of the
light curve shows that the emission increased 
faster initially, it had a well-defined local maximum at $t \approx 10$\,ks, and then it growed 
again, but more gradually this time, until the end of the observation. At the very end, there is a hint 
that the emission was stopping to increase. Such a long rise phase suggests that we 
observed an uninterrupted sequence of flare events involving an extended coronal region. 
However, the faster initial rise and the local emission \mbox{peak -- coupled} to the earlier hardness 
ratio \mbox{peak -- resembles} the evolution of a self-standing flaring episode possibly occurring
(at least during these initial stages) in a
single loop
%system 
\citep[e.g.][]{rea07}, whose decay merged with the ignition of the later 
overlapping events. 
%Therefore, this initial event is analogous to the kind of flares observed for 
%instance on CC Eri \citep{cre07} with the only difference that here we are less able to
%resolve it as separated. 
%The first peak at $t = 10-11$ ks in the light curve 
Fig.~\ref{fig2.1} demonstrates that this peak
is indeed significative and 
%it 
cannot be attributed to the effect of noise in the light curve. 
%In Fig.~\ref{fig2.1}, we plot the cumulative 
This figure shows the
cumulative distribution of counts for 200 simulated constantly increasing 
light curves\footnote{Constantly increasing light curves have been
chosen for the simulations since the global enhancement mostly shows
a linear pattern.}
(for $t \ge 5$\,ks) with Poissonian noise (shadowed region)
%. The cumulative distribution of counts in the observed light curve is
%over-plotted as a continuous line. 
together with the cumulative distribution of counts observed during the
rise (continuous line).
%The plot shows that the simulations cannot 
Clearly, the simulations do not
reproduce the observed distribution of counts for the first peak in the
light curve. The differences are high enough to justify the treatment of 
%the first peak of the light curve 
this event as independent.
In the following, we work under this hypothesis.

%The flare rise ($t \ge 5$\,ks) 
%was therefore split into two sections 
%to be analyzed
%separately: the first part ($\sim 10$~ks long), associated to the ignition of a single loop, and the 
%second one ($\sim 25$~ks long), associated to a more complex evolution that 
%can be described 
%as a two-ribbon flare (in agreement with the spectral analysis results; see
%$\S$~\ref{sec:specanalysis}).

%________________________________________________ Figure
%______________________________________________________

The large number of counts collected from the source in the EPIC
cameras allowed us to divide
the observation in several time intervals with enough signal to perform a reliable
spectral analysis {($\ge 1000$ counts in the quiescent and intervals 1 and 2, and
$\ge 2000$ counts in the remaining intervals after background subtraction\footnote{{Intervals
1 and 2 have been studied separately to investigate the evolution of the parameters of the 
flaring plasma in its very first phases.}})}. 
This permitted: 1) to investigate the nature of the processes taking place in our target;
and 2) to derive physical properties of it.

\section{Spectral analysis}
\label{sec:specanalysis}

%After a careful analysis using different divisions, we finally decided 
%to split the observation 
For the spectral analysis, the observation was split
into 10 time-intervals (vertical
dashed lines in Fig.~\ref{fig1}). These intervals
sample different features in the light curve.
%, such as the first increase {\bf (starting $5$\,ks after the beginning)} 
%and the following brief decrease.
We used the XSPEC spectral fitting package \citep{arn96, arn04} in the
PN, MOS1, and MOS2 detectors simultaneously. We adopted the 
%Astrophysical Plasma Emission Database (APED), 
%which contains the relevant atomic data for both continuum and line emission
%\citep{smi01}, included in the XSPEC software.
Astrophysical Plasma Emission Code \citep[APEC,][]{smi01} 
included in the XSPEC software. APEC calculates spectral models
for hot, optically thin plasmas using 
the Astrophysical Plasma Emission Database
\citep[APED,][]{2001ASPC..247..161S}, that contains the relevant 
atomic data for calculating both the continuum and line emission.
Interstellar absorption was taken 
into account using the interstellar photo-electric absorption cross-sections 
of \citet{mor83}, also available in XSPEC.

%\subsection{General X-ray properties in the quiescent state}

\subsection{Quiescent state}

\begin{figure}[!t]
\includegraphics[bb= 68 355 480 600, width=8.0cm,clip=true]{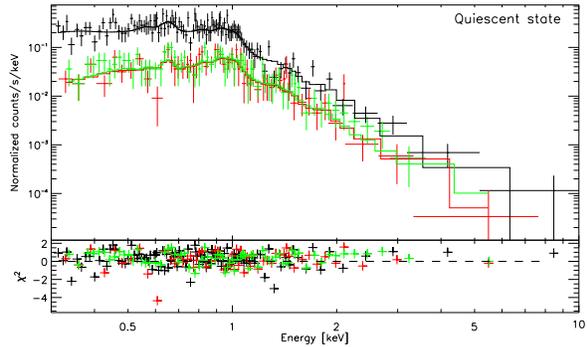}
\caption{Observed EPIC PN and MOS spectra of TWA 11B during the 
quiescent state (marked with Q in Fig.~\ref{fig1}).}
\label{fig3}
\end{figure}

%In this work, 
The lowest (constant) count-rate level, found at the beginning of the
observation, is assumed to be the quiescent state. The X-ray luminosity of
the star during this lapse of time {(assuming the same distance than the 
primary $d = 67 \pm 3$ pc)} is $\log L_\mathrm{X} \mathrm{[erg\,s^{-1}]} = 29.35$, 
corresponding to $\log (L_\mathrm{X}/L_\mathrm{bol}) = -3.1$. 
These values are typical of both pre-main sequence stars members 
of young stellar associations \citep[e.g.][]{kas03} and older 
field M dwarfs showing high X-ray activity \citep{rob05, cre07}.
The best fit to the X-ray quiescent spectrum (Fig.~\ref{fig3}) is given
by a plasma model with two equally-weighted thermal components
%, having the same weight 
%(see Table~\ref{tab_quiescent} 
%for a summary of the results of the fit). 
(the output parameters of this fit are summarized
in Table~\ref{tab_quiescent}).
{Although high uncertainties are obtained for the hydrogen column 
density in our fit, the value of $N_\mathrm{H}$ obtained by us is very similar
to that determined for other members of the TW Hya Association 
\citep[e.g.][]{ste04,arg05}. Besides, the interstellar extinction was 
found to be negligible in the general direction of TW Hya 
\citep{ruc83}.}

%________________________________________________ Figure
%______________________________________________________

%_____________________________________________________________ Table
\begin{table}[!t]
\caption[]{Output parameters from fitting the X-ray EPIC spectra 
          of the quiescent state with a $2T$-model.}
\label{tab_quiescent}
\centering
\begin{tabular}{lcl}
%\hline\hline
\noalign{\smallskip}
\noalign{\smallskip}
\hline
\noalign{\smallskip}
$NH$ & = & $1.7^{+2.6}_{-1.7} \times 10^{20}$ cm$^{-2}$ \\
\noalign{\smallskip}
$Z$ & = & $0.17^{+0.08}_{-0.06}$ $Z_\odot$ \\
\noalign{\smallskip}
$kT_1$ & = & $0.27^{+0.03}_{-0.03}$ keV \\
\noalign{\smallskip}
$EM_1$ & = & $2.2^{+1.4}_{-0.8} \times 10^{52}$ cm$^{-3}$ \\
\noalign{\smallskip}
$kT_2$ & = & $0.98^{+0.08}_{-0.08}$ keV \\
\noalign{\smallskip}
$EM_2$ & = & $2.2^{+0.6}_{-0.5} \times 10^{52}$ cm$^{-3}$ \\
\noalign{\smallskip}
$\chi^2_{\rm red}$ (d.o.f.) & = & $0.86 \ (176)$ \\
\noalign{\smallskip}
\hline
\end{tabular}
\end{table}
%___________________________________________________________________

%\subsection{Evolution of the temperature during the flare}
%\label{sec:spec_evolution}

%________________________________________________ Figure
%\begin{figure}[!t]
%\includegraphics[width=8.0cm,clip=true]{f4.ps}
%\caption{Density-temperature diagram
%%Emission measure and temperature evolution 
%during the flare. $EM_F^{1/2}$ is used as a proxy for the 
%density. Emission measures are given in units of $10^{52}$ cm$^{-3}$ and 
%temperatures in Kelvin.}
%\label{fig4}
%\end{figure}
%______________________________________________________

\subsection{Flaring state}
\label{sec:flaring}

%For the spectral analysis, 
Synthesized stellar-like spectra
of solar flares (previously subtracted by the quiescent
``background'' spectrum) are generally \mbox{well-fitted} with
a single thermal component \citep[1-$T$ model, see][]{2001ApJ...557..906R}.
Consequently, to analyze physical properties of the flaring plasma 
in our observation,
we first subtracted the spectrum of the quiescent state from
the observed spectrum in each time-interval. Then we fitted a
$1T$-model, leaving its temperature ($kT_\mathrm{F}$) and 
emission measure ($EM_\mathrm{F}$) as free parameters.
The values of the column density ($NH$) and abundance ($Z$) 
were fixed to those derived for
the quiescent state. With this technique, we {interpret} that
the coronal spectrum during the flare results from adding to the quiescent spectrum
a third thermal component, which is ascribed to heated material filling
the flaring loops \citep{2001ApJ...557..906R, cre07}. %formed through reconnections.
Models with additional temperature components were checked, but did not
improve the fit results significantly.

Results from spectral fitting are given in Table~\ref{tab2}.
As expected from the hardness ratio evolution (Fig.~\ref{fig2}), the temperature 
peaked in time-segment 1. %during the first time-interval.
Then, it decreased gradually but almost continuously until the end of the observation. 
On the other hand, the emission measure increased continuosly from the beginning 
of the rise and starts to decrease only near the end of the observation.
%If we look at this evolution of the temperature and emission as a whole of the 
%conductive cooling phase of a single-loop flare 
%\citep[see a detailed description in][]{rea07}, where the heat pulse stops and the 
%plasma starts to cool while still filling the loop from below.

%_____________________________________________________________ Table
\begin{table*}[!t]
\centering
\caption[]{Spectral results for the flaring component in each time-segment of the
light curve (excepting the quiescent time-interval).
%. The quiescent time-interval is omitted.
%Results of the spectral fitting in each time-interval from the first enhancement of the light curve.
}
\label{tab2}
\begin{tabular}{c c c c c c c c}
\noalign{\smallskip}
\hline\hline
\noalign{\smallskip}
  Time-segment & Time-interval & Central time & $kT_\mathrm{F}$ & $EM_\mathrm{F}$ & $\chi^2_{\rm red}$ (d.o.f.)\\
         & (ks)          & (ks)         & (keV)  & ($10^{52}$ cm$^{-3}$) & \\
\noalign{\smallskip}
\hline
\noalign{\smallskip}
1 & 5 -- 8   & 6.5  & 8$^{+60}_{-4}$         & 0.72$^{+0.20}_{-0.17}$ & 1.05 (129) \\
\noalign{\smallskip}
2 & 8 -- 11  & 9.5  & 2.4$^{+0.4}_{-0.3}$    & 4.0$^{+0.3}_{-0.3}$    & 0.87 (215) \\
\noalign{\smallskip}
3 & 11 -- 17 & 14   & 2.5$^{+0.4}_{-0.3}$    & 3.88$^{+0.27}_{-0.28}$ & 1.08 (369) \\
\noalign{\smallskip}
4 & 17 -- 22 & 19.5 & 2.27$^{+0.25}_{-0.27}$ & 5.3$^{+0.3}_{-0.3}$    & 0.87 (355) \\
\noalign{\smallskip}
5 & 22 -- 27 & 24.5 & 1.88$^{+0.14}_{-0.14}$ & 7.2$^{+0.3}_{-0.3}$    & 0.89 (406) \\
\noalign{\smallskip}
6 & 27 -- 32 & 29.5 & 1.66$^{+0.18}_{-0.08}$ & 8.3$^{+0.3}_{-0.3}$    & 1.03 (430) \\
\noalign{\smallskip}
7 & 32 -- 36 & 34   & 1.31$^{+0.05}_{-0.05}$ & 9.2$^{+0.4}_{-0.4}$    & 0.97 (379) \\
\noalign{\smallskip}
8 & 36 -- 38 & 37   & 1.22$^{+0.06}_{-0.06}$ & 9.9$^{+0.5}_{-0.5}$    & 1.01 (243) \\
\noalign{\smallskip}
9 & 38 -- 41 & 39.5 & 0.99$^{+0.04}_{-0.04}$ & 9.7$^{+0.4}_{-0.4}$    & 0.86 (313) \\
\noalign{\smallskip}
\hline
\end{tabular}
\end{table*}
%___________________________________________________________________

\subsubsection{Evolution of the emission measure distribution}
\label{sec:EMDs}

\begin{figure}[!t]
\includegraphics[width=8.2cm,clip=true]{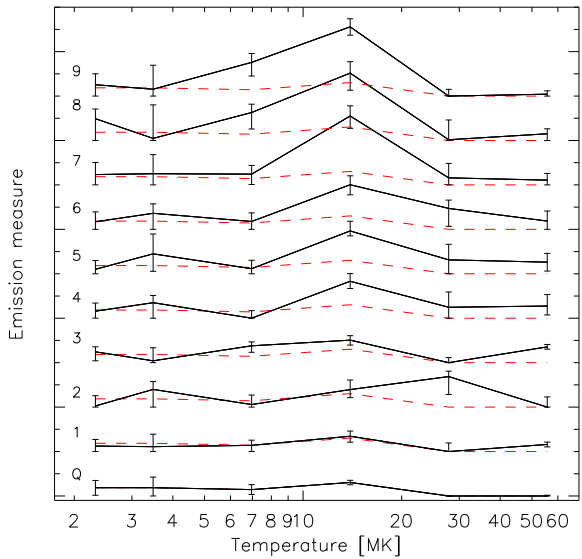}
\caption{EMDs as derived with the $6T$-model.
%Emission measure of the six thermal components used for the fit.
The curves are
shifted in the y-axis to better observe the flare evolution in time. 
Each one of the EMDs (shown as continuous lines) is labeled with the same number as its
corresponding time-segment (see Table~\ref{tab2}). The quiescent is marked as Q. 
For comparison, the EMD of the quiescent state is also overplotted as a red, 
dashed line at the same zero level than each one of the rest EMDs.}
\label{fig4}
\end{figure}

In order to approximate physically more realistic continuous emission measure
distributions (EMDs) of the plasma, we used
%To investigate the evolution with time of the emission measure, we decided to use 
a multi-temperature model as in \citet{rob05}. In that work, the authors 
used a $6T$-model on a logarithmic, almost equidistant grid with temperatures 
fixed at 0.2, 0.3, 0.6, 1.2, 2.4, and 4.8 keV (which correspond to 2.3, 3.5, 
7.0, 14.0, 28.0 and 56.0 MK), sampling those spectral regions where 
the XMM-Newton detectors are more sensitive.
%simulating a coronal temperature gradient. 
%In the fitting, the abundance and column density
To fit our spectra, the values of $NH$ and $Z$ 
were fixed to those previously determined for the quiescent state.
%As the six temperatures are also fixed, 
Thus, the only variables 
%in our model
are the emission measures. 
%of each one of the six thermal components. 
Note that in this case we are fitting each spectrum as a whole
(i.e., the quiescent spectrum was not subtracted from the rest
of the spectra).
%Our 
Results from applying this $6T$-model to our observation 
are shown in Table~\ref{tab2.1} and Fig.~\ref{fig4}, and are 
summarized in the following items:

%_____________________________________________________________ Table
\begin{table*}[!t]
\centering
\caption[]{Output parameters (emission measures, in units of $10^{52}$ cm$^{-3}$) 
           from fitting the whole X-ray EPIC spectra to a $6T$-model with
           temperatures fixed at 0.2, 0.3, 0.6, 1.2, 2.4 and 4.8\,keV.
           $EM'_1$ is the emission measure corresponding to the thermal 
           component of 0.2\,keV, $EM'_2$ is that of 0.3\,keV, and so on.
           Uncertainties are calculated for a 2.7$\sigma$ confidence level.}
\label{tab2.1}
%\small
\begin{tabular}{ccccccc}
\noalign{\smallskip}
\hline\hline
\noalign{\smallskip}
Time Interval & $EM'_1$ & $EM'_2$ & $EM'_3$ & $EM'_4$ & $EM'_5$ & $EM'_6$ \\
\noalign{\smallskip}
\hline
\noalign{\smallskip}
Quiescent & $1.00_{-0.91}^{+0.88}$ & $1.00_{-1.00}^{+1.28}$ & $0.77_{-0.59}^{+0.61}$ & $1.63_{-0.28}^{+0.26}$ & $0.00_{-0.00}^{+......}$ & $0.00_{-0.00}^{+0.09}$ \\
\noalign{\smallskip}
1 & $0.66_{-0.66}^{+0.80}$ & $0.59_{-0.59}^{+1.51}$ & $0.75_{-0.75}^{+0.63}$ & $1.85_{-0.72}^{+0.63}$ & $0.00_{-0.00}^{+1.04}$ & $0.85_{-0.30}^{+0.29}$ \\
\noalign{\smallskip}
2 & $0.13_{-0.13}^{+1.24}$ & $2.14_{-2.14}^{+0.96}$ & $0.31_{-0.31}^{+1.15}$ & $2.12_{-0.97}^{+1.15}$ & $3.70_{-2.17}^{+0.66}$ & $0.00_{-0.00}^{+1.23}$ \\
\noalign{\smallskip}
3 & $1.31_{-1.09}^{+0.61}$ & $0.22_{-0.22}^{+1.58}$ & $2.04_{-0.78}^{+0.48}$ & $2.73_{-0.60}^{+0.54}$ & $0.00_{-0.00}^{+0.61}$ & $1.91_{-0.28}^{+0.29}$ \\
\noalign{\smallskip}
4 & $0.85_{-0.84}^{+0.98}$ & $1.88_{-1.88}^{+0.88}$ & $0.01_{-0.01}^{+0.92}$ & $4.48_{-0.88}^{+0.92}$ & $1.32_{-1.32}^{+1.86}$ & $1.48_{-1.08}^{+1.39}$ \\
\noalign{\smallskip}
5 & $0.54_{-0.54}^{+1.07}$ & $2.42_{-2.11}^{+2.40}$ & $0.62_{-0.62}^{+1.02}$ & $5.19_{-0.60}^{+1.17}$ & $1.69_{-1.69}^{+1.89}$ & $1.41_{-1.07}^{+1.06}$ \\
\noalign{\smallskip}
6 & $0.92_{-0.92}^{+1.20}$ & $1.94_{-1.94}^{+1.14}$ & $0.96_{-0.96}^{+1.04}$ & $5.41_{-1.22}^{+1.08}$ & $2.53_{-2.17}^{+1.01}$ & $1.01_{-1.01}^{+1.22}$ \\
\noalign{\smallskip}
7 & $1.26_{-1.24}^{+1.46}$ & $1.36_{-1.36}^{+2.32}$ & $1.30_{-1.25}^{+1.06}$ & $8.35_{-1.49}^{+1.24}$ & $0.86_{-0.86}^{+1.75}$ & $0.58_{-0.58}^{+0.82}$ \\
\noalign{\smallskip}
8 & $2.64_{-2.64}^{+1.17}$ & $0.25_{-0.25}^{+4.07}$ & $3.39_{-2.00}^{+1.01}$ & $8.17_{-2.07}^{+1.38}$ & $0.09_{-0.09}^{+2.40}$ & $0.81_{-0.81}^{+0.60}$ \\
\noalign{\smallskip}
9 & $1.38_{-1.38}^{+1.32}$ & $0.83_{-0.83}^{+2.90}$ & $4.08_{-1.66}^{+1.08}$ & $8.39_{-1.02}^{+0.98}$ & $0.00_{-0.00}^{+0.82}$ & $0.23_{-0.23}^{+0.40}$ \\
\noalign{\smallskip}
\hline
\end{tabular}
\end{table*}
%___________________________________________________________________

\begin{itemize}

\item The amount of plasma emitting at temperatures above 28\,MK is negligible
in the quiescent state.
%The quiescent spectrum (`Q' in the figure) is well described by a 4T-model.
%During the first time-segment of the flare (where the maximum temperature was
%observed when three thermal components where used; see Table~\ref{tab2}) the
%emission measure of the sixth component ($kT_6 \sim 5$~keV) becomes significant
%(note that each error in Table~\ref{tab2.1} is at 2.7$\sigma$ level) while the rest of
%the spectrum remains as in the quiescent state.

\item Plasma emitting at the highest temperatures ($\sim$ 56\,MK)
appears at the beginning of the first (faster) rise (time-segment 1),
while the rest of the EMD curve remains as in the quiescent state. 
Note that in this time-segment we obtained the maximum temperature for the
flaring component when fitting it with the $1T$-model (see first part of
$\S$~\ref{sec:flaring} and Table~\ref{tab2}).

\item The only significant difference between the EMD of the 
quiescent state and that derived for time-segment 2 
(where the light curve of the first flare peaks) is an excess of 
plasma emitting at temperatures
around 28\,MK. We interpret it as the cooling of the material
at higher temperatures that was detected in time-segment 1
together with additional plasma evaporated after
the temperature peak was reached in time-segment 1 (note that 
the total emission measure of the flaring component in time-segment 2 
is higher than that measured in time-segment 1, as it can also be seen
 in Table~\ref{tab2}).
%During time-segment 2 (where the light curve peaked) the sixth thermal
%component disappears ($EM_6 = 0$) while $EM_5$ (corresponding to
%$kT_5 \sim 2.5$~keV) increases above the quiescent value.
%The rest of the spectrum is still as in the quiescent state
%(taking errors into account).

\item The excess of plasma emitting in the region around 28\,MK
that was observed in time-segment 2 continues cooling towards lower temperatures
during time-segment 3. At the same time, a high quantity of plasma appears
again at the highest temperatures ($\sim$ 56\,MK). This quantity is
even larger than that measured in time-segment 1 and is approximately
coincident with the beginning of the second (more gradual) rise. 
%(which, as seen in $\S$~\ref{sec:observations}, is probably associated to 
%a two-ribbon flare).
%%In time-segment 3, the emission measure of the fifth component ($EM_5$) 
%%becomes again as in the quiescent while thermal components 3 and 4 show higher 
%%emission measures. As in time-segment 1, at this time a peak in $EM_6$, corresponding 
%%to the higher thermal component, is observed.

\item During time-segments 4 -- 7, the amount of plasma emitting
at the highest temperatures remains still high, while 
a large excess of material (compared to the quiescent state) appears 
also between 7 and 28\,MK (reaching the maximum always at a temperature 
$\approx$~14\,MK). The existence of intense, sustained heating is 
needed to explain the maintenance of plasma at temperatures around 56\,MK.
The evolution of the rest of the EMD curve can be interpreted as the
cooling of the continuously new appearing very hot plasma 
together with heated material that evaporates from lower 
layers and fills the flaring loops.
%During time-segments 4--8, the value of $EM_6$ decreases continuously, while 
%the emission measure of the first thermal components increases. At the end of the
%observation (time-segments 8 and 9), the emission measure of the hard thermal 
%components is not significant (taking errors into account). 

\item During time-segments 8 and 9 the emission at temperatures 
$\gsim 28$\,MK turns back to the level found at the quiescent
state, which is an indication of the strong heating not being present
any more. However, the height of the maximum in the EMD curve is even larger
than in the previous time-segments, appearing also excess emission at 
even lower temperatures (down to $\sim 3$\,MK).

\end{itemize}

%_____________________________________________________________ Figure
%__________________________________________________________________

%\textbf{
%These results support the idea, suggested by us in $\S$~\ref{sec:observations}, 
%of the observed light curve being the consequence of two separate flare events: 
%the first one starting at time-segment 1 and the 
%second one beginning at time-segment 3, during the decay phase of the first flare. 
%The first event could be ascribed to a single loop or a small group of loops
%probably dominated by one of them, 
%since its ignition occurs only at time-segment 1 (as suggested by the plasma
%appearing at very high temperatures in its corresponding EMD). 
%On the other hand, the evolution of the whole EMD during the second event 
%evidences continuous heating of material (decreasing in efficiency) probably 
%occurring during continuous reconnections, as it happens in a two-ribbon flare. 
%}

\section{Flare modeling}
\label{sec:model}

Results from $\S$~\ref{sec:observations} and $\S$~\ref{sec:specanalysis}
support the idea of the observed rise ($t \ge 5$\,ks)
being the consequence of two separate flare events: the first more
impulsive one (event A) starting in time-segment 1, followed by a
more gradual energy release (event B) having
a beginning that is merged with the decay phase of event A and becoming
dominant at the end of time-segment 3. But, what kind of magnetic
structures are producing each one of these events? According to
\cite{rea07}, multiple loop structures can be involved in a flare, but
this frequently occurs only in its late phases. The initial phases of an
X-ray flare are usually quite localized and one can reasonably assume the
presence of a single dominant loop \citep[e.g.][]{2001SoPh..204...91A,rea04}.
This assumption is realistic enough in most of the
observed rise phases, where the impulsive heating typically involves a
dominant loop
(while later residual heating may be released in other similar adjacent
loops). The clear evidence of a delay between the temperature peak and the
density peak is consistent with a single loop model, at least until the
moment in which this maximum density is reached. This delay is often
observed both in solar flares \citep[e.g.][]{1993A&A...267..586S},
and in stellar flares
\citep[e.g.][]{1988A&A...205..181V,1989A&A...213..245V,fav00,2000A&A...356..627M,2002A&A...392..585S},
even in very long ones \citep{fav05,2008ApJ...688..418G}. The presence of this
delay is indeed a signature of a relatively short heat pulse (the larger
the delay, the shorter the heat pulse is) and of the coherent
hydrodynamic evolution of plasma confined in a loop. Arcades and
two-ribbon flares
are instead characterized by strong and/or lasting heating \citep{kop84}
and/or irregular light curves \citep{2001SoPh..204...91A,rea04}.
Thus, event A is likely characterized (or dominated) by a
single flaring loop since its ignition occurs only at time-segment 1 (as
suggested by the plasma appearing at very high temperatures in its
corresponding EMD, see $\S$~\ref{sec:EMDs}) and shows a delay between the
temperature peak and the density peak reached by the flaring plasma in it
(see Table~\ref{tab2}). Event B is probably triggered by event A. The evolution of
the whole EMD during event B
evidences continuous, but decreasing in efficiency, heating of material
and is consistent with event B being a two-ribbon flare. In $\S$~\ref{sec52} we
demonstrate that the light curve of event B is indeed well-reproduced
by a two-ribbon flare model. Continual reconnections in an arcade may
also be
responsible for event B. However, in this case
%the observed light curve would probably show a more irregular pattern
one would expect a more irregular pattern in the light curve
than that observed in Fig.~\ref{fig1},
and the EMD evolution would not necessarily point to a kind of heating
that is decreasing in efficiency with time, as we have observed.

\subsection{Event A: the single loop flare}
\label{sec:singleloop}

We used the results from fitting the spectra with the $1T$-model 
(see $\S$~\ref{sec:specanalysis}), given in Table~\ref{tab2}, to 
determine the length of the loop involved in event A.
For the temperature in time-segment 1, we used $kT = 4$~keV  
(the lower limit from the fit) since its upper limit is undetermined 
(see Table~\ref{tab2}).

For this event, the complete rise phase is observed and instead we 
miss satisfactory information from the decay phase. A complete 
theoretical analysis of the rise phase of a flare occurring 
in a single loop is given by \citet{rea07}. The author 
showed that the loop half-length ($L$) can be determined from 
the maximum temperature reached by the flaring plasma ($T_0$)
%($T_0 = 0.13 T^{1.16}_\mathrm{obs}$ for the EPIC 
%instrument, where $T_\mathrm{obs}$ is the temperature
%measured for the flaring component by spectral fitting)
and the temperature and time in which the density peaks
($T_\mathrm{M}$ and $t_\mathrm{M}$, respectively) by: 

%\begin{equation}
%L_9 \approx 2.5 \frac{\psi^2}{\ln \psi}  T^{1/2}_{0,7} \Delta t_\mathrm{0-M,3}
%\label{eq2}
%\end{equation}
%
%being $\psi = T_0/T_\mathrm{M}$, or equivalently:

\begin{equation}
%L_9 \approx 3 \psi^2 T^{1/2}_{0,7} t_\mathrm{M,3} 
%L \approx 3 \cdot 10^{5/2} \left(\frac{T_0}{T_\mathrm{M}}\right) ^2 T^{1/2}_{0} t_\mathrm{M}
L \approx 3 \cdot 10^{5/2} t_\mathrm{M} \frac{T_0^{5/2}}{T_\mathrm{M}^{2}} 
\label{eq1}
\end{equation}

\noindent where
%$\psi = T_0/T_\mathrm{M}$ and
all the parameters are given in c.g.s. units. The maximum 
temperature ($T_0$) is related to the temperature measured for the 
flaring component by spectral fitting ($T_\mathrm{obs}$). For the 
EPIC instrument:
%through the expression:}

\begin{equation}
T_0 = 0.13 T^{1.16}_\mathrm{obs}
\label{eqT0}
\end{equation}

For event A, the maximum temperature and maximum emission measure were 
reached in the time-segments 1 and 2, respectively. For a single loop description, the 
(square root of the) emission measure becomes a good proxy of the density, because 
the loop volume presumably does not change much during the event.
In this case, from Eq.~\ref{eq1} we obtain a loop half-length
$L = 1.8 \pm 0.3 \times 10^{11}$~cm, i.e. $\sim 4 \pm 1$ $R_\star$ 
\citep[assuming a stellar radius $R_\star = 0.64$ $R_\odot$ -- from the pre-main sequence 
models of][for an M2.5 star with 8 Myr, such as TWA 11B is supposed to be]{sie00}.
%assuming 
%a stellar radius $R_\star = 0.64$ $R_\odot$ -- from the pre-main sequence models
%of \citet{sie00} for an M2.5 star with 8 Myr, such as TWA 11B 
%is supposed to be). 
Although this value is relatively large, similar loop sizes have already been
derived for young stars in star-forming regions \citep{fav05,fra07}.
%the Orion Nebula Complex \citep{fav05}. 
As far as we are concerned, this would be
the first time that such a relatively long loop is detected in a 
star older than 3 -- 4\,Myr. 
%members of the Taurus star-forming region.
%Nevertheless, the obtained value for the loop half-length 
%should be taken as an upper limit. (perche` questa frase?)

This loop would fill a volume
$V \approx 2.3 \times 10^{31}$ cm$^3$ \citep[see][for further details 
on the relations used for determining this quantity]{rea07}. 
Assuming a semi-circular geometry, its aspect ($r/L$,
with $r$ being the loop cross-section) would be of the order of 2~\%,
which is quite lower than that observed in 
the Sun ($\sim 10$~\%), but compatible with the results in \citet{fav05}. 
In Table~\ref{tab3} we summarize the main parameters of the loop
involved in event A, i.e. those already mentioned and maximum density
at the loop apex ($n_\mathrm{M}$), average density in the loop
when the maximum density is reached at the apex ($n_\mathrm{avg}$)
and cross-section area of the loop ($A$). Such parameters were 
determined with the relations given in \citet{rea07}. 

Comparing our results 
with those obtained by \citet{rea07} for events observed in Algol, AB~Dor, and 
Prox~Cen, we conclude that: 
1) our densities seem to be similar to those determined in the Algol
and Prox Cen flares;
and 2) the loop volume and its subtended area are of the same order of magnitude as
those found for Prox Cen, another M star. 
Note that the largeness of some of the errors shown in Table~\ref{tab3} is
a consequence of the propagation of errors in the equations.

%_____________________________________________________________ Table
\begin{table}[!t]
\caption[]{
Parameters of the flaring loop involved in event A, derived as in \citet{rea07}.
We have assumed a semi-circular geometry.}
\label{tab3}
\centering
\begin{tabular}{llc}
\noalign{\smallskip}
\hline\hline
\noalign{\smallskip}
  Parameter & \multicolumn{1}{c}{Units} & Value \\
\noalign{\smallskip}
\hline
\noalign{\smallskip}
%$n_0$                         & $10^{10}$ cm$^{-3}$ & $7.1 \pm 1.2$ \\
%\noalign{\smallskip}
$n_\mathrm{M}$       & $10^{10}$ cm$^{-3}$ & $2.0 \pm 0.7$ \\ 
\noalign{\smallskip}
$n_\mathrm{avg}$    & $10^{10}$ cm$^{-3}$ & $4.1 \pm 0.7$ \\ 
\noalign{\smallskip}
$V$                                  & $10^{31}$ cm$^{3}$  & $2.3 \pm 0.8$ \\
\noalign{\smallskip}
$A$                                  & $10^{20}$ cm$^{2}$  & $0.6 \pm 0.2$ \\
\noalign{\smallskip}
$r$                                   & $10^{10}$ cm              & $0.4 \pm 0.1$ \\
\noalign{\smallskip}
$L$                                  & $10^{11}$ cm              & $1.8 \pm 0.3$ \\
\noalign{\smallskip}
\hline
\noalign{\smallskip}
\end{tabular}
\end{table}
%___________________________________________________________________

For the sake of completeness, we made the exercise to repeat the same analysis to 
the entire flare event (A$+$B) as if it all occurred in a single loop. For a flare temperature 
peaking in time-segment 1 ($t_0 \approx 6.5$ ks), and the density peaking in 
time-segment 8 ($t_\mathrm{M} \approx 37$ ks), using Eq.~\ref{eq1}, we obtained 
$L = 1.3 \times 10^{13}$ cm, i.e. $L \sim 260 R_\star$ . Obviously, such a long loop would 
be easily destroyed by the stellar rotation. Clearly, this result makes no sense.

\subsection{Event B: the two-ribbon flare}
\label{sec52}

For the study of event B, we used the two-ribbon flare model 
by \citet{kop84} extended to the stellar case \citep{pol88}.
This model supposes that a disruptive event opens a loop arcade,
being the open field lines then driven toward a radial neutral sheet
(above the magnetic neutral line) where they reconnect at progressively
higher altitudes. Thus, the continuous heating provided by these reconnections
is capable of reproducing the temporal profile of the energy rate released during both 
the rise and decay phases of a two-ribbon flare. By analogy to that
observed on the Sun, the model considers that the arcade of loops
is extended along the East-West direction (i.e., axial symmetry around
the polar axis is assumed). It also assumes that the magnetic field is
potential between the stellar surface and the location of 
the neutral line and extends radially outwards from there.
The magnetic field in the meridional planes of the arcade can therefore be expressed
in terms of a single lobe of a Legendre polynomial of degree $n$. 
Note that: 
\renewcommand{\labelenumi}{\roman{enumi}}
\begin{enumerate}
\item Each lobe is latitudinally bounded by radial magnetic fields.
\item The arcade corresponds to one lobe axisymmetrically continued over some
longitude in the East-West direction.
\item Through an appropriate choice of $n$, one can find a lobe placed in
the range of latitudes covered by the active region. However, as spatial
information is available only for the Sun, 
%our flaring region is 
stellar flaring regions are generally assumed to
be centered on the equator for odd $n$ and to end at the equator for even $n$.
\item As stellar observations cannot provide any information on the time-dependent
rise of the neutral point, it is assumed to mimic the solar case. 
Thus, it follows an exponential law of the form given by Eq.~\ref{eq:neutralpoint},
where $y$ is the height of the neutral point (in units of $R_\star$, measured
from the star's center), $t$ is the time (measured from the beginning of the
two-ribbon flare, $t_\mathrm{ini}$), $t_0$ is a time-constant, and $H_\mathrm{m}$ is the maximum height 
reached by the reconnection point during its upward movement (measured from the
star's surface). $H_\mathrm{m}$
is typically chosen to be equal to the latitudinal extent of the arcade,
which is in turn linked with $n$ (see Eqs.~\ref{eq:Hmn1} and~\ref{eq:Hmn2}).
\end{enumerate} 

%\begin{equation}
%\label{eq:neutralpoint}
%   y = 1 + \frac{H_\mathrm{m}}{R_\star} (1 - e^{-\frac{t}{t_0}})
%\end{equation}
\begin{align}
\label{eq:neutralpoint}
y  &= 1 + \frac{H_\mathrm{m}}{R_\star} (1 - e^{-\frac{t}{t_0}})\\
\label{eq:Hmn1}
H_\mathrm{m} & \approx \frac{\pi}{n + 1/2} R_\star & \text{for}~n > 2\\
\label{eq:Hmn2}
H_\mathrm{m} & \approx \frac{\pi}{2} R_\star & \text{for}~n = 2
\end{align}

Under all these assumptions, the rate of magnetic energy released
by the reconnecting arcade per radian of longitude ($dE/dt$) can be
expressed as:

\begin{eqnarray}
\label{eq:dEdt}
\nonumber \frac{dE}{dt}  = \frac{1}{8\pi} 2n (n+1) (2n+1)^2 R^3_\star B^2_\mathrm{m}~~~~~~~~~\\
         \times \frac{I_{1,2}(n)}{P^2_n(\theta_{1,2})} 
         \frac{y^{2n} [y^{2n+1} - 1]}{[n + (n+1) y^{2n+1}]^3}
         \left( \frac{dy}{dt} \right)
\end{eqnarray}

where $I_{1,2}(n) = \int P^2_n(\theta) \textrm{d}(\cos \theta)$
evaluated between the latitudinal borders of the lobe, 
$P_n(\theta)$ is the Legendre polynomial of degree $n$,
$\theta$ is the co-latitude, $\theta_{1,2}$ is the co-latitude
of either boundary of the lobe, and $B_\mathrm{m}$ is the maximum
surface magnetic field in the active region. 
%The factors $B_\mathrm{m}^2$ and $I_{1,2}(n)$ merely define 
The factor $B_\mathrm{m}^2$ merely defines
the normalization of the energy release light curve, while $t_0$ and $n$ determine
its shape. As Eq.~\ref{eq:dEdt} is given per radian of longitude, a length ($l$) 
must be assumed for the arcade in order to calculate its total 
energy-release rate. Solar two-ribbon flares occur in loop arcades whose length is
typically about 1.5 times their width ($l \approx 1.5H_\mathrm{m}$). 
In our study, we adopted this ratio for any given $n$. 

Since, in the view of many authors, the initiation process for the flare 
itself might be the result of a more rapid (nearly explosive) reconnection than
the reconnection process about which \citet{kop84} and \citet{pol88} speak,
they stressed that their model is applicable only after the initial flare
trigger mechanism is terminated.
In fact, the model seems not to be able to describe the time of impulsive heating 
%As stressed by \citet{kop84} and \citet{pol88}, their model is applicable 
%only after the initial flare trigger mechanism is terminated since
%the initiation process for the flare itself might be the result of a more rapid 
%(nearly explosive) reconnection than the reconnection process about which they 
%speak.
%% although \citet{pneu82} suggested that reconnection may 
%% start in the earliest phase of loop structure development. 
%In particular, it seems not to be able to model the time of impulsive heating 
and steeply increasing temperatures. However, the impulsive phase typically 
ceases before reaching the $50~\%$ flare peak level 
in the soft X-ray bandpass commonly used for stellar observations. 
At this time in our observations, the temperature is gradually decreasing. 
Hence, the model is applicable from time-segment 4 of the light curve
to the end of our observations (time-segment 3 is rejected also 
for avoiding possible contamination from the decay of event A).

In order to apply the described model, the photospheric magnetic field
in the flaring region (from which $B_\mathrm{m}$ can be 
determinated), and its latitudinal location and size (which dictate $n$) should 
be known. In the solar case, observations provide all these data, whereas
in the stellar case, at best, they can only be inferred indirectly.
Since we are unaware of the location, size, and magnetic field strength of
the active region that we are studying, we treated $n$ and $B_\mathrm{m}$ as free
parameters to be determined from the best fit of the model to the observations.

We created a grid of values for the free parameters $n$, $t_0$, 
and $t_\mathrm{ini}$ to fit event B with the two-ribbon flare model
(the lower limit of $t_\mathrm{ini}$ was fixed at the beginning 
of the whole enhancement
%observed rise 
because of obvious physical reasons).
For each set of these parameters, we determined the $B_\mathrm{m}$ that best fit
the data by minimizing the $\chi^2$ value, which is defined as:

\begin{equation}
  \chi^2 = \sum_{i=1}^N \left(\frac{L_{\mathrm{mod},i} - L_{\mathrm{obs},i}}{\Delta L_{\mathrm{obs},i}}\right)^2
\end{equation}

\noindent where $L_\mathrm{mod}$ is the expected luminosity from the model
($L_\mathrm{mod} = f \cdot q \cdot l \cdot dE/dt$), $L_\mathrm{obs}$
is the observed luminosity, $\Delta L_\mathrm{obs}$ is the error in the observed 
luminosity, and $N$ the number of time-intervals with which we fitted the two-ribbon model
(from time-segment 4 to 9).
Following the solar 
analogy \citep[][see $\S$~\ref{sec:int}]{can80}, we assumed
the measured radiative losses in the X-ray band to be $\approx 10~\%$
of the thermal energy generated as consequence of magnetic reconnections,
%global flare energy losses, 
that is $f \approx 0.1$ \citep[see][for details]{pol88}.
%which traduces in using a multiplicative 
%factor ($f \approx 0.1$) in Eq.~\ref{eq:dEdt} \citep[see][for details]{pol88}.
%%($f \approx 0.1$).}
Actually,
only a fraction ($q < 1$) of the liberated magnetic energy is
indeed used to heat the plasma (thermal energy that is subsequently lost via radiation and
conduction), while the rest is
%lost mainly under the form of
transformed into mechanical energy, into fast particles ejected
from the corona, etc. 
At this point, we want to notice that we did not find a unique solution corresponding
to a single set of parameters, but a number of solutions producing a good fit
($\chi^2 \sim 1$).

\begin{figure}[!t]
%\label{fig1}
\includegraphics[width=8.3cm,clip=true]{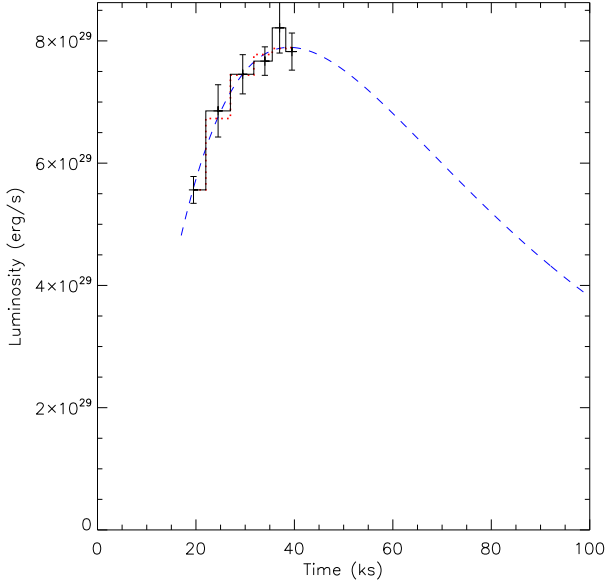}
\caption{Best fit to the observations using the two-ribbon model
for $n = 2$ (see Table~\ref{tab4}). Note that each set of parameters 
in Table~\ref{tab4} produces similar results in the fitted region while having
different slopes in the previous part and in the decay.
The light curve segments used to fit the two-ribbon model (time-segments 4--9, see 
also Fig.~\ref{fig1}) is shown 
with a solid line. The dashed line is the model. We have also plotted the 
average values of the model in each time-interval during event B (dotted line) 
for a clearer comparison between the observed light curve and the model.
\label{fig5}}
\end{figure}

In Table~\ref{tab4}, we show the two-ribbon flare parameters for some of 
the good fits we have obtained.
We plot these results together with the observations in Fig.~\ref{fig5}.
%our results for 4 values of the 
%%polynomial degree ($n$)
%degree of the Legendre polynomial. 
%Although 
Other good fits were found also for larger values of $n$
(i.e., active regions with smaller width and, therefore, shorter loops),
%we find quite improbable a scenario in which a long loop, such as that 
%%ignited at the beginning of the observation, 
%involved in event A (see $\S$~\ref{sec:singleloop}), triggers a 
%subsequent flare (event B) in 
%loops at much lower heights. 
but note that
a smaller active region needs higher
%High loops need also less
surface magnetic fields for reproducing a given energy-release rate.
%to confine the evaporated material below the height of the neutral point.
Thus, loop systems that reach larger altitudes --~i.e., small values of $n$~--
%Small values of $n$ --~i.e., high loop systems~--
may be more realistic in our case, although other loop configurations 
cannot be excluded (see $\S$~\ref{s:fluor}).
%Note also that
%a wider active region needs smaller
%%High loops need also less 
%surface magnetic fields for reproducing a given energy release rate. 
%%to confine the evaporated material below the height of the neutral point. 

%________________________________________________ Figure
%______________________________________________________

%_____________________________________________________________ Table
\begin{table}[!h]
\scriptsize
\caption[]{Two-ribbon flare parameters resulting from fitting the model to the 
observed energy release rate from event B.}
%observed luminosity of TWA~11B.}
\label{tab4}
\centering
\begin{tabular}{lcccc}
%\hline\hline
\noalign{\smallskip}
\noalign{\smallskip}
\hline
\noalign{\smallskip}
Polynomial degree                  & 2 & 3 & 5 & 10 \\
\noalign{\smallskip}
Region width [deg]                   &  90$^\circ$ & 53$^\circ$ & 33$^\circ$ & 17$^\circ$ \\
\noalign{\smallskip}
$H_\mathrm{m}$ [R$_\star$]  &  1.57  & 0.90 & 0.57 & 0.30 \\
\noalign{\smallskip}
$L^\dag_\mathrm{m}$ [R$_\star$]   &  2.46 & 1.41 & 0.89 & 0.47 \\
\noalign{\smallskip}
$t_\mathrm{ini}$ [ks]                 &  7.0   & 6.5   & 6.0   &  5.5   \\
\noalign{\smallskip}
$t_0$ [ks]                                    &  247  & 192  &   183  & 174  \\
\noalign{\smallskip}
$B_\mathrm{m} \sqrt{q}$ [G]   &  440  & 730  & 1050  & 1830 \\
\noalign{\smallskip}
v$_\mathrm{rise}$ [km s$^{-1}$] & 2.8 & 2.1 & 1.4 & 0.8 \\
\noalign{\smallskip}
$N_\mathrm{e}$ [$\times 10^{10}$ cm$^{-3}$]  & 1.1 & 2.6 & 5.1 & 13.3 \\
\noalign{\smallskip}
$\chi^2$                                      & 1.02 & 1.02 & 1.04 & 1.04 \\
\noalign{\smallskip}
\hline
\noalign{\smallskip}
\end{tabular}

\scriptsize{$\dag$ $L$ is the loop semi-length 
determined from the maximum height $H_\mathrm{m}$
assuming a semi-circular geometry.}
\end{table}
%___________________________________________________________________

The values given in Table~\ref{tab4} for the maximum surface magnetic field 
%%($B_\mathrm{m}$)
were determined assuming that all the magnetic energy is used to heat the 
plasma that fills the flaring loops ($q=1$). 
%Actually, 
%only a fraction ($q < 1$) of the liberated magnetic energy is 
%indeed used to heat the plasma (energy that is subsequently lost via radiation and
%conduction), while the rest is
%%lost mainly under the form of 
%transformed into mechanical energy, into fast particles ejected
%from the corona, etc. 
\citet{kop84} used $q = 0.003$ for a solar flare. 
On the other hand, 
during the analysis of a stellar flare, \citet{gue99} found acceptable solutions with $q < 0.05$
for $n = 2$ and $q \approx 0.01-0.02$ for $n = 3-4$. For event B
%the flare we are analyzing, 
we found that small values of $q$ ($\sim 0.01-0.02$) 
%give 
imply
very strong 
photospheric magnetic fields in the flaring region
($B_\mathrm{m}$ $\approx 3 - 18$ kG). 
%($B_\mathrm{m} \approx 4 - 12$ kG). 
Using $q = 0.05$, we obtained $B_\mathrm{m}$ $\approx 2$~kG for $n = 2$ and 
%$3.2-4.7$~kG for $n = 3-5$. %In the table, we give the value $B_\mathrm{m} \sqrt{q}$.
$B_\mathrm{m}$ $\approx 3 - 8$~kG for $n = 3 - 10$.

%, which is directly related with the maximum height of
%the two-ribbon systems (see Eq.~\ref{eq:Hm}). 
%
%
%$B_\mathrm{m} \sqrt(q)$ is the normalization factor...
%
%The maximum magnetic field
%$B_\mathrm{m}$ is given in terms of the parameter $q$, that gives the fraction of
%liberated magnetic energy used to heat the plasma.}

\section{The fluorescent Fe 6.4 keV line}
\label{s:fluor}

\begin{figure*}[!t]
\includegraphics[width=8.0cm,clip=true]{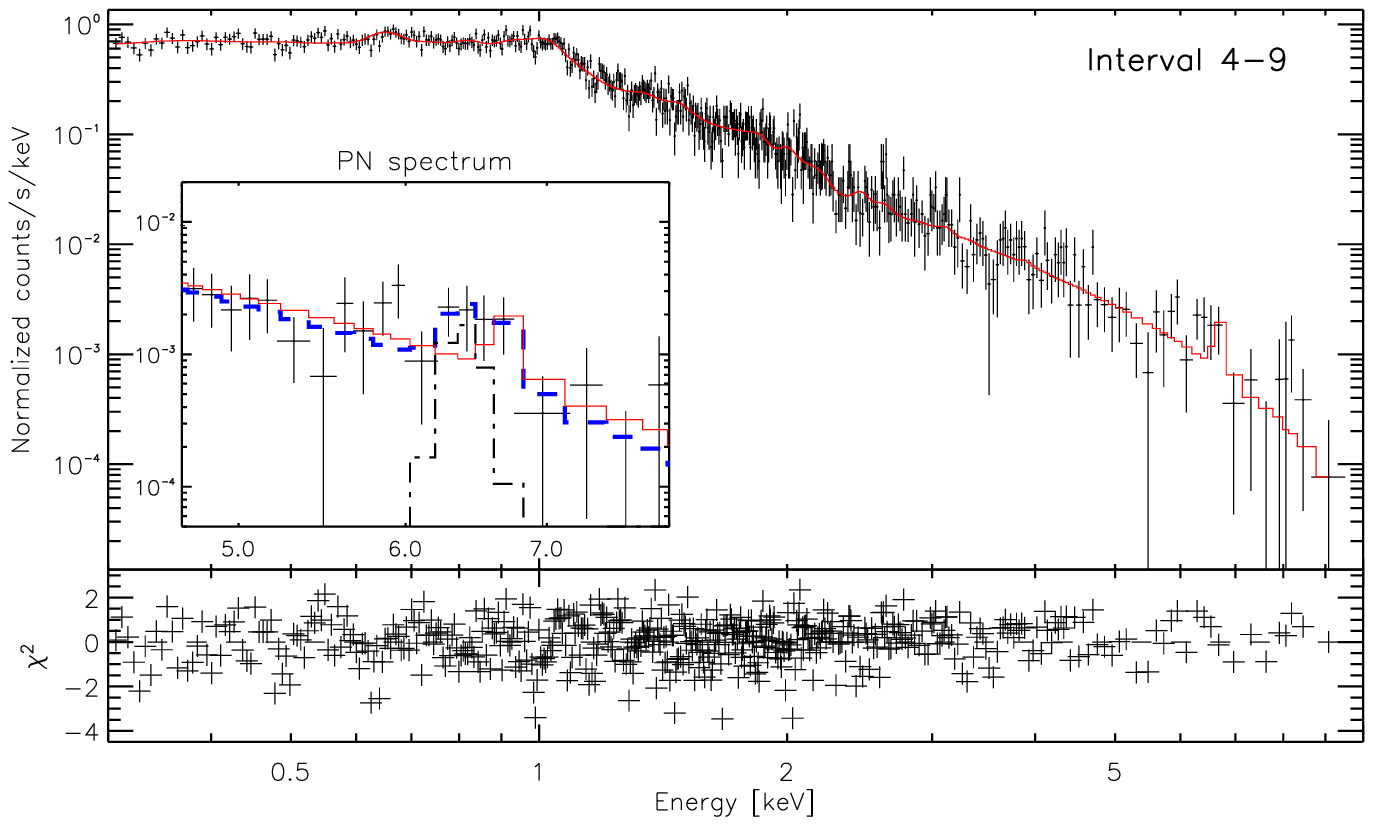}
\includegraphics[width=8.0cm,clip=true]{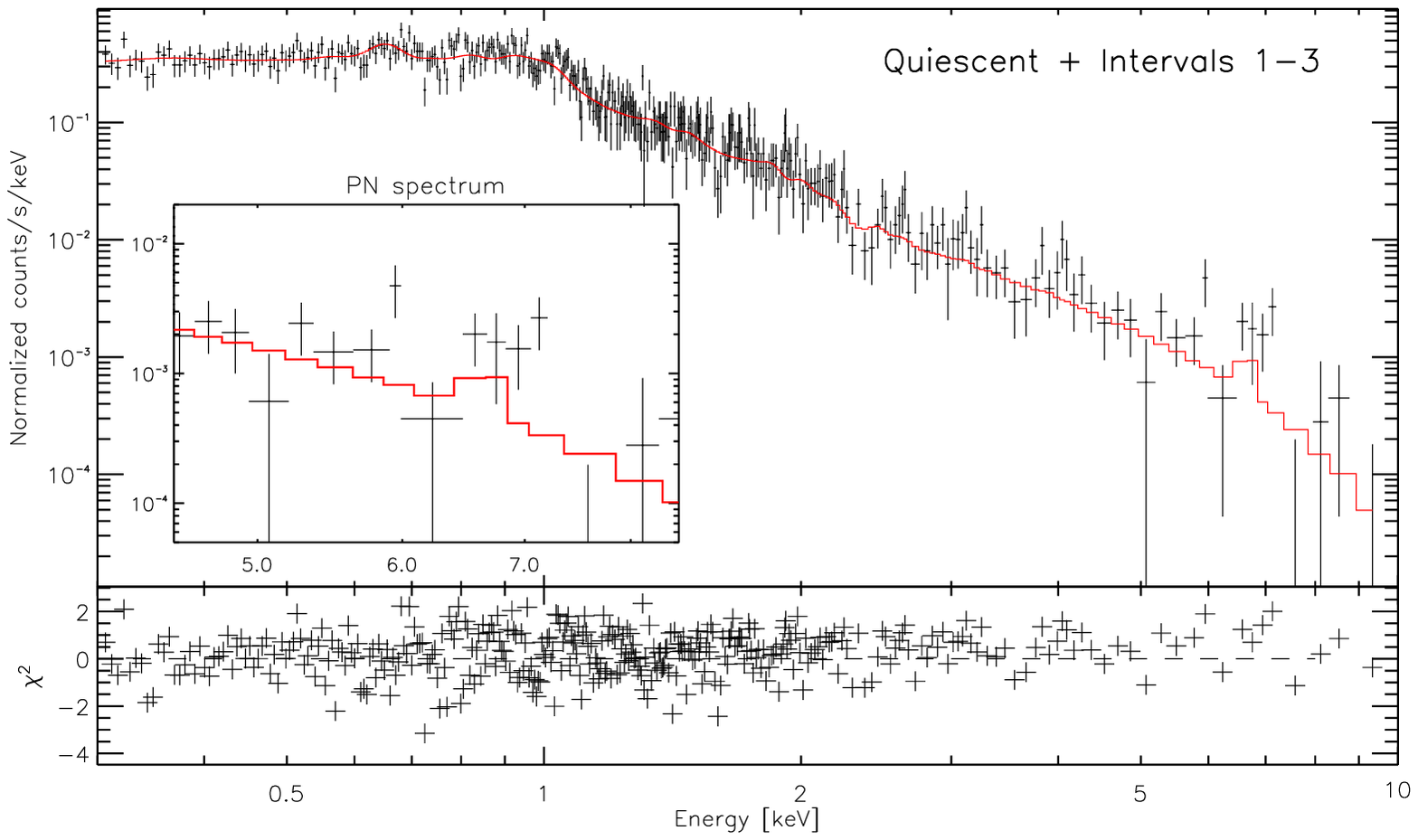}
\caption{Left: PN spectrum of TWA 11B integrated in the 
time-segments 4 -- 9, with the clear excess emission in 6.4~keV. The (red) continuous line 
is the fitted 3$T$-model. In the small window, the (blue) dashed-line is the 
fitted 3$T$-model + Gaussian component. The Gaussian 
component is also plotted 
%obtained from the fitting is over-plotted 
as a dotted-dashed line. Right: Same as figure at 
the left but in the time-segments 1 -- 3 plus quiescent. Here, the excess 
emission in 6.4~keV is clearly not present.}
\label{fig6}
\end{figure*}

During the inspection of the X-ray spectrum in the different time-intervals, we observed 
a feature (excess emission not reproduced by the plasma model) close to the Fe K$_\alpha$ 
line at redder wavelengths.
We identified this feature as the Fe fluorescent line at 6.4~keV. 
It was noticeable only during the time-segments 4 to 9, while it was 
neither in time-segments 1 to 3 nor in the quiescent state 
(see Fig.~\ref{fig6}). Unfortunately, we could not monitor
the time-variation of this emission line since the individual spectra
of each time-interval have no counts enough for accurately fitting the line.
%To quantify this emission line, we performed a fit to the 
%spectrum integrated from time-segments 4 to 9 (inclusive) with 
%an interstellar absorbed 3$T$ plasma model and an 
%additional Gaussian line component at 6.4~keV. 
Thus, to quantify the excess emission at 6.4~keV in our observations, 
we performed spectral fitting
using an interstellar absorbed 3$T$-plasma model and an
additional Gaussian line component at 6.4~keV. We fitted, on
the one hand, the spectrum integrated from the quiescent to
time-segment 3 (inclusive) and, on the other hand, the
spectrum integrated from time-segment 4 to 9 (inclusive).
The gaussian flux obtained in this way for the former spectrum is zero.
%Therefore, hereafter we only discuss the results found for the
%latter spectrum. 
%
The best fit results are shown in Table~\ref{tab5}.  
Note that we left free only the temperature and the emission measure
of the third (hottest) thermal component and the gaussian, while the 
rest of the parameters were fixed to the values obtained for the quiescent 
state (Table~\ref{tab_quiescent}).

%________________________________________________ Figure
%______________________________________________________

%_____________________________________________________________ Table
\begin{table}[!t]
%\footnotesize
\small
\centering
\caption[]{Best fitting values
%Results of the fitting 
(3$T$ plasma model + Gaussian) for the spectrum integrated
in the time-segments 4 -- 9. Only the temperature and the emission measure of
%values for 
the third thermal component and the Gaussian are 
listed because the remaining parameters were fixed to the values estimated
for the quiescent state. The given $\chi^2_{\rm red}$ and d.o.f. values 
refer to the total fit.} 
\label{tab5}
\centering
\begin{tabular}{lcl}
%\hline\hline
\noalign{\smallskip}
\noalign{\smallskip}
\hline
%\noalign{\smallskip}
%$NH$ & = & $1.7^{+2.6}_{-1.7} \times 10^{20}$ cm$^{-2}$ \\
%\noalign{\smallskip}
%$Z$ & = & $0.17^{+0.08}_{-0.06}$ $Z_\odot$ \\
%\noalign{\smallskip}
%$KT_1$ & = & $0.27^{+0.03}_{-0.03}$ keV \\
%\noalign{\smallskip}
%$EM_1$ & = & $2.2^{+1.4}_{-0.8} \times 10^{52}$ cm$^{-3}$ \\
%\noalign{\smallskip}
%$KT_2$ & = & $0.98^{+0.08}_{-0.08}$ keV \\
%\noalign{\smallskip}
%$EM_2$ & = & $2.2^{+0.6}_{-0.5} \times 10^{52}$ cm$^{-3}$ \\
\noalign{\smallskip}
$kT_3$ (keV) & = & $2.02^{+0.13}_{-0.12}$ \\
\noalign{\smallskip}
$EM_3$ ($10^{52}$ cm$^{-3}$) & = & $4.7^{+0.3}_{-0.3}$ \\
\noalign{\smallskip}
Gaussian central energy (keV) & = & $6.4^{+0.9}_{-0.6}$ \\
\noalign{\smallskip}
Gaussian flux ($10^{-6}$ ph\,cm$^{-2}$\,s$^{-1}$) & = & $0.9^{+0.7}_{-1.3}$ \\
\noalign{\smallskip}
Gaussian $\sigma$ (eV) & = & $9.7^{+0.6}_{-9.7}$ \\
\noalign{\smallskip}
$\chi^2_{\rm red}$ [d.o.f.] & = & $0.96 \ [499]$ \\
\noalign{\smallskip}
\hline
\end{tabular}
\end{table}
%___________________________________________________________________

In the past, 
fluorescent Fe emission 
%has been observed commonly 
were commonly observed in classical T Tauri stars and protostars 
\citep{tsu05, fav05, gia07, sci08}, where it has been attributed to the incidence 
of X-ray emission onto the proto-planetary disk or into the circumstellar gas surrounding 
very young objects. However, the Fe 6.4~keV line has been observed also in the giant 
star HR~9024 \citep{tes08} and the RS~CVn system II~Peg \citep{ost07}. In the case of 
HR~9024, the authors attributed its presence to the incidence of hard X-rays onto 
the photosphere, in concordance with what is observed in the Sun. In II~Peg, the 
excitation mechanism was ascribed to electron impact ionization of photospheric Fe. 

In our observation, we have not statistics enough to perform a robust analysis of 
the fluorescent line. Nevertheless, some constraints can be given. On the one hand, the 
line is not observed during the first event. This may be consistent with the scenario of the 
long loop for this event. The efficiency of the fluorescence decreases with the distance 
of the X-ray source to the photosphere \citep[e.g.][]{dra08}. Therefore, if the fluoresce line
were produced by photoionization, it should not be observed during flares occurring in 
long loops. On the other hand, the 
presence of the line during the second event could be indicative of a not very high 
loop system, if it were produced by photoionization. In contrast, if the line were
produced by collisional ionization, there would be none constraint to the loop 
height. A result in favor with the non-photoionization nature of the fluorescence
line is shown in \citet{cze07}. In their study, the authors modeled the illuminating
input spectrum and obtained line fluxes below the observations. With our data, 
the collisional production of the fluorescence line cannot be discarded.

\section{Final remarks and conclusions}

In this paper, we analyzed the long rise phase of a flare observed in an XMM-Newton 
archive data of the $\sim 8$~Myr old star TWA~11B. 
%Based on a number of 
%results obtained from time-resolved spectral fitting and the inspection of the 
%light curve and the hardness-ratio curve, we concluded that the star suffered 
%a first ignition of a single loop that subsequently triggered a two-ribbon 
%flare.
The analysis of the light curve, of the hardness-ratio curve, and of the 
time-resolved spectra consistently indicates that probably the flare first involved 
mainly a single loop and then propagated to a loop arcade becoming a proper 
two-ribbon flare. 
%The observation was split into two phases, the first one dominated by the 
%single-loop flare and the second one dominated by the two-ribbon system.
%We analyzed each phase separately using models of \citet{rea07} and 
%\citet{kop84}, respectively. 
We split our analysis into three parts: the quiescent state, the single-loop flare 
(event A), and the two-ribbon system (event B). Event A was studied with the 
analysis described in \citet{rea07}. For event B, we used the stellar version of the
\citet{kop84}'s solar two-ribbon flare model that is able to provide 
some limited information about the late flaring structures.

For the single loop, we obtained a semi-length of 
approximately  $1.8 \pm 0.3 \times 10^{11}$~cm ($\sim 4 \pm 1$~R$_\star$), with 
a volume $V = 2.3 \times 10^{31}$~cm$^3$ and a cross-section $r/L \sim 2\%$.
These values are comparable to those found by \citet{fav05} in Orion members.
This fact makes us suggest that large and thin loops are common in young
active stars.
For the two-ribbon system, different results consistent with the observed data
%solutions 
were found. Good fits ($\chi^2 \sim 1$) 
were obtained for both small and large values of $n$ (i.e. for high and 
%low height systems
short loop arcades). Bearing the semi-length of the 
first loop in mind, the more realistic
scenario for the two-ribbon system is that with long loops. In any case,
the estimated values of the maximum surface magnetic field in the 
flaring region result to be 
%are 
quite strong, reaching 2 -- 8 kG 
%for the case in which 
when a large fraction 
of the liberated magnetic energy ($q = 0.05$) is 
%used 
assumed to heat the plasma, and up to 18 kG when 
$q \sim 0.01$. 

During the inspection of the X-ray spectrum,
%in different time-intervals, 
we observed the Fe fluorescent line at 6.4~keV 
%in
during the time in which the two-ribbon system evolved
(time-segments $4-9$). 
%From the equivalent width of this line and the angle
%subtended by the reflector (determined using the maximum height of the loop),
%we determined a column density $N_\mathrm{H} \sim 2.5 \times 10^{25}$~cm$^{-2}$,
%which is very high. Since photons should experience significant
%Compton scattering at 
%%those densities
%such a high column density, we suggest that the fluorescence 
%was produced by collisional ionization of the photosphere instead 
%of photoionization. 
%
The absence of the line during the first event may be
consistent with the long loop scenario for event A. In contrast, the 
detection of the fluorescence line during the second event could be 
indicative of a not very high loop system involved in the two ribbon
flare, if the line were produced by photoionization. Otherwise, 
there would be none constraint to the loop system height.
%As far as we are concerned, this is only the third clear 
%detection of Fe photospheric fluorescence
%{\bf in stars other than the Sun}.

\acknowledgments

JLS acknowledges financial support by the 
PRICIT project S-0505/ESP-0237 (ASTROCAM) of the Comunidad 
Aut\'onoma de Madrid (Spain). ICC acknowledges
support from the Spanish {\it Ministerio de
Educaci\'on y Ciencia}, under the grant F.P.U. \mbox{AP2001-0475};
and from the Marie Curie Actions grant MERG-CT-2007-046535.
The Madrid group acknowledges partial support 
by the Programa Nacional de Astronom\'{\i}a y Astrof\'{\i}sica of the Spanish 
Ministerio de Educaci\'on y Ciencia (MEC), under grants AYA2008-00695 
and AYA2008-06423-C03-03. FR acknowledges support from Italian 
Ministero dell'Universit\`a e della Ricerca (MIUR). 
GM acknowledges support from the Agenzia Spaziale Italiana (ASI) and 
the Istituto Nazionale di Astrofisica (INAF) under grant ASI-INAF I/088/06/0. 
{We would like to thank the referee and the editor for useful comments 
and discussion that helped us to improve this manuscript.}

\end{document}